\documentclass[10pt,journal,twocolumn]{article}
\usepackage{amsmath,amsfonts}
\usepackage{algorithmic}
\usepackage{array}
\usepackage{textcomp}
\usepackage{stfloats}
\usepackage{url}
\usepackage{verbatim}
\usepackage{graphicx}
\def\BibTeX{{\rm B\kern-.05em{\sc i\kern-.025em b}\kern-.08em
    T\kern-.1667em\lower.7ex\hbox{E}\kern-.125emX}}
\usepackage{balance}
\usepackage{multirow}
\usepackage{pgf}
\usepackage{cite}
\usepackage{subcaption}
\usepackage{makecell}
\usepackage{enumitem}

\usepackage{booktabs}
\usepackage{orcidlink}
\usepackage{siunitx}
\usepackage{color, colortbl}
\usepackage{import}
\usepackage{svg}
\definecolor{LightCyan}{rgb}{0.88,1,1}
\definecolor{LightYellow}{rgb}{1, 1, 0.8}

\def\traningrepos{5,414 }
\def\totalrepos{6,766 }
\def\totalfiles{2,430,138 }

\usepackage{seqsplit}
\newcommand{\ttbreak}[1]{\texttt{\seqsplit{#1}}}

\newcommand{\revision}[1]{{\color{black}#1}}
\newcommand{\minor}[1]{{\color{black}#1}}

\begin{document}

\title{ConcernBERT: Learning Responsibilities Using Class Membership}

\author{Jason~Lefever~\orcidlink{0000-0002-9505-265X}, 
Jiahao~Xu~\orcidlink{0009-0001-3282-0033}, Yuanfang~Cai~\orcidlink{0000-0002-2690-8557}, Rick~Kazman~\orcidlink{0000-0003-0392-2783}, and~Ernst~Pisch~\orcidlink{0009-0007-4083-763X}%
}

\date{}

\markboth{IEEE Transactions on Software Engineering}{Lefever \MakeLowercase{\textit{et al.}}: ConcernBERT: Learning Responsibilities Using Class Membership}

\maketitle

\begin{abstract}  
The principles of \textit{separation of concerns}, \textit{high cohesion}, and \textit{single responsibility} are among the most well-known in software design. However, their application often remains philosophical rather than actionable, relying heavily on developers' intuition and experience. Many software tasks, such as god class decomposition, extract class refactoring, and cohesion measurement, depend on techniques for identifying cohesive groups of program entities, that is, entities that collectively fulfill a common responsibility. Yet reliably identifying such groups remains a challenge. \revision{In this paper, we propose ConcernBERT, a BERT-based embedding model trained at the entity level that uses triplet loss to directly optimize the relative positioning of methods and attributes in the embedding space, and uses class-membership context to learn responsibilities and concerns. We also contribute a large-scale replication dataset for training and evaluation}. Our dataset spans over two million Java files across more than six thousand repositories. To evaluate ConcernBERT, we merge methods from two or more classes into unlabeled groups and test the model's ability to recover the original class memberships. ConcernBERT achieves significantly higher performance than existing models, demonstrating its effectiveness at encoding concern-level semantics and establishing a strong foundation for downstream tasks such as architecture recovery, extract class refactoring, and cohesion measurement.
\end{abstract}

\section{Introduction}
\label{sec:introduction}
The principle of ''separation of concerns'' was first introduced in 1982~\cite{Dijkstra1982}, with broader discussions of modularity dating back to at least the early 1970s~\cite{Parnas1972}. While this principle has shaped decades of thinking about software design, its application in practice still largely depends on the developer's intuition and experience. Identifying program entities that serve the same responsibility or concern is a fundamental step in many canonical software engineering tasks, such as architectural recovery~\cite{Maletic1999,Maletic2000,Maletic2001,Marcus2004,Kuhn2005,Kuhn2007,Baldi2008,Garcia2011,Sajnani2012,Shahbazian2018,Zhang2023}, extract class refactoring~\cite{Lucia2008,Bavota2010,Bavota2011,Bavota2013,Bavota2014,Akash2019}, and quality metrics for coupling~\cite{Poshyvanyk2008,Ujhazi2010,Gethers2010} and cohesion~\cite{Marcus2005,Ujhazi2010,Liu2009,Chen2017,Miholca2021}.  

\revision{Researchers have long recognized that identifying "single responsibility" is a complex problem involving multiple factors~\cite{marinescu2004:metric-based-rule, Lee2012ASA}. Beyond semantic similarity, prior work on god class decomposition and cohesion measurement also considers dependency relationships among methods and attributes, as well as both internal and external interactions of a class.} As an important factor in "single responsibility," state-of-the-art approaches increasingly leverage language models to compare program entities based on semantic similarity. Prevailing approaches consider entities more "semantically similar" when they contain similar sets of tokens.

Although such methods have shown promise, it remains unclear to what extent token-based semantic similarity can truly reflect the underlying concerns or responsibilities of the program entities. 

\begin{figure*}[htbp]
    \centering
    \setlength{\fboxsep}{8pt}  
    \setlength{\fboxrule}{0.1pt}  

    \makebox[\textwidth]{
        \begin{subfigure}[b]{0.45\textwidth}
            \centering
            \fbox{\includegraphics[scale=1]{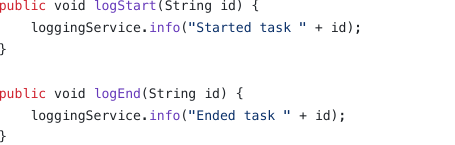}}
            \caption{Same Concern / Same Tokens}
            \label{fig:same-concern-similar-text}
        \end{subfigure}
        \hspace{1em}
        \begin{subfigure}[b]{0.45\textwidth}
            \centering
            \fbox{\includegraphics[scale=1]{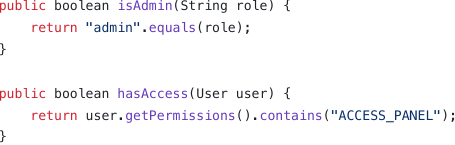}}
            \caption{Same Concern / Different Tokens}
            \label{fig:same-concern-dissimilar-text}
        \end{subfigure}
    }

    \vspace{1.5em}

    \makebox[\textwidth]{
        \begin{subfigure}[b]{0.45\textwidth}
            \centering
            \fbox{\includegraphics[scale=1]{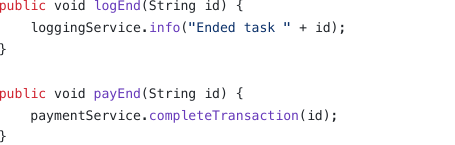}}
            \caption{Different Concern / Same Tokens}
            \label{fig:different-concern-similar-text}
        \end{subfigure}
        \hspace{1em}
        \begin{subfigure}[b]{0.45\textwidth}
            \centering
            \fbox{\includegraphics[scale=1]{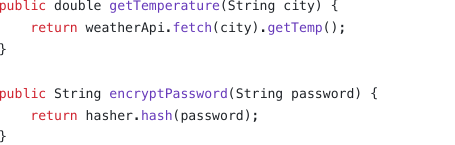}}
            \caption{Different Concern / Different Tokens}
            \label{fig:different-concern-dissimilar-text}
        \end{subfigure}
    }

    \caption{The token distributions of program entities are not necessarily aligned with their concern.}
    \label{fig:four-panel-example}
\end{figure*}

In Figure~\ref{fig:four-panel-example}, we use four examples to illustrate why token similarity alone is insufficient to capture the true concerns of program entities. In (a), the two logging methods share both vocabulary and concern, so token similarity suffices. In (b), the two access control methods address the same concern but use entirely different tokens, demonstrating that semantically related methods may not share vocabulary. Conversely, (c) presents two methods with \revision{similar tokens} but serving unrelated concerns---one logs events, while the other handles payment---showing that token similarity can be misleading. Finally, (d) shows unrelated concerns with different tokens. While in (a) and (d), concern similarity and token similarity align, this is not the case in (b) and (c), demonstrating that token similarity does not consistently reflect semantic responsibility, motivating the need for models that capture deeper concern-level relationships.

To overcome the limitations of token-based similarity, we introduce \emph{ConcernBERT}, a novel BERT-based model that embeds program entities, such as methods and attributes, into a high-dimensional latent space where entities with shared responsibilities are positioned close together. The key insight behind ConcernBERT is that class membership itself provides a supervision signal for learning concern-level similarity: developers group methods into classes based on shared responsibility, and this grouping decision encodes semantic information that pure token analysis cannot capture. ConcernBERT is trained to position members of the same class closer to each other than to members of other classes. By learning from millions of such grouping decisions across thousands of repositories, ConcernBERT captures patterns of concern-level similarity that generalize to unseen code, even when entities do not share overlapping tokens. This training paradigm enables the model to capture deeper concern-level relationships that are not apparent from surface-level lexical similarity, an essential capability for identifying the responsibilities of program entities.

ConcernBERT \revision{achieves} this through \revision{two key innovations in its training process:
\begin{enumerate}
\item \textit{Entity-level supervision:} Unlike existing models that use token-level masked learning objectives, ConcernBERT is trained at the entity level using triplet loss, directly optimizing the relative positioning of methods and attributes in the embedding space according to their concern.

\item \textit{Class membership as context:} Positive examples are drawn from members of the same Java structure (class, interface, or enum), treating developer-assigned class membership as an implicit label for shared responsibility. This assumption---that classes encapsulate a single concern---transforms millions of existing design decisions into training signal.
\end{enumerate}
}
\revision{Another contribution is our replication package and the full dataset used to train and evaluate ConcernBERT, containing a total of \totalfiles source files across \totalrepos repositories~\cite{Lefever2025a}.}
Through this process, ConcernBERT captures deeper concern-level semantics that lexical or token-based models cannot capture. 

We evaluate ConcernBERT's ability to capture program concerns using repeated trials of the \emph{Class-Membership Recovery Test}. \revision{In each trial, methods from 2, 10, 50, or 100 classes are merged into an unlabeled group, embedded, and clustered.} The resulting clusters are compared against the original class labels. A model that accurately recovers the original class structure is considered better at distinguishing program concerns. This evaluation design closely reflects real-world software development tasks such as extract class refactoring~\cite{Lucia2008,Fokaefs2012,Bavota2013,Bavota2014,Akash2019,Alzahrani2021,Alzahrani2022,Lefever2025}.

We compare ConcernBERT against seven models: CodeBERT~\cite{Feng2020}, \revision{GraphCodeBERT~\cite{Guo2021}, CodeT5+~\cite{Wang2023}, ModernBERT~\cite{Warner2024}}, Latent Semantic Indexing (LSI)~\cite{Deerwester1990}, Latent Dirichlet Allocation (LDA)~\cite{Blei2003}, and Doc2Vec~\cite{Le2014}. Our evaluation includes 88,027 small-group trials (merging two classes) and 1,283 large-group trials (merging 100 classes), spanning over one million methods across 176,560 classes. Because different models perform best under different clustering configurations, we evaluate each across twelve configurations combining four clustering algorithms with three distance metrics, then report each model's optimal configuration and compare ConcernBERT's performance against these optimal baselines.

Our results show that ConcernBERT substantially outperforms all reference models across both small and large groups. On small-group tests, it achieves a 31\% improvement over the next-best model LSI. On the more challenging large-group tests, ConcernBERT achieves a 55\% improvement. Compared to CodeBERT, which shares its architecture but not its training process, ConcernBERT shows a 115\% gain on small-group tests and 94\% on large-group tests. All differences are statistically significant ($p < 0.01$), demonstrating that ConcernBERT's class membership-based training produces embeddings better aligned with actual responsibilities than existing models.

These results suggest that ConcernBERT provides a stronger foundation for tasks such as architectural recovery, extract class refactoring, and coupling and cohesion metrics. By learning similarity directly from class membership rather than surface-level token overlap, ConcernBERT moves beyond traditional approaches and delivers more reliable insights into program responsibilities, with potential to translate philosophical principles like ``\emph{high cohesion}" and ``\emph{single-responsibility}" into operable metrics and tooling for more principled, automated software engineering practices. \minor{ConcernBERT itself is not intended as an end-to-end refactoring or architecture recovery tool; rather, it serves as a foundational embedding component to support these downstream tasks.}
\section{Related Work}\label{sec:related-work}

In this section, we review embedding models commonly used in tasks such as extract class refactoring, and coupling or cohesion metrics, and the baseline models we chose to compare against ConcernBERT. Originally developed in natural language processing for documents like books or articles, these models generate high-dimensional vectors from text. In software tasks, these ``documents'' are program entities such as functions, classes, or files.

\revision{
\textit{CodeBERT.} CodeBERT~\cite{Feng2020} and related encoder-only models such as~\cite{cubert, huggingface_codeberta} are finding increasing use within both software engineering practice and research~\cite{Mosel2023,Keim2020,Keim2020a}. These models advance language modeling and document representation primarily due to their attention mechanisms and large-scale pretraining. They are typically trained using masked language modeling, in which the model learns to ``fill in the blank'' given surrounding context. However, like other models with token-level training objectives, this results in program entities being considered similar based on the surface similarity of their lexical content, which does not necessarily indicate that they address the same concern.

\textit{GraphCodeBERT.} GraphCodeBERT~\cite{graphcodebert} and related models~\cite{unixcoder, wang2022syncobert, GraphCode2Vec} are pre-trained models that explicitly incorporate the structural information of source code. In addition to token sequences, GraphCodeBERT leverages data-flow graphs to model semantic relationships such as variable definitions and uses, enabling the model to capture both syntactic and semantic dependencies in code. By jointly learning from code tokens and their associated data-flow structures, these models produce more structure-aware representations and have been shown to improve performance on a variety of code understanding tasks, including code search, clone detection, and code refinement.

\textit{CodeT5+.} CodeT5+~\cite{codet5+} is a large-scale encoder-decoder model for source code understanding and generation that extends the original CodeT5~\cite{wang2021codet5} framework with improved pre-training objectives, larger model sizes, and more diverse training data. Built on the T5 architecture, CodeT5+ is pre-trained on both unimodal code data and bimodal code (natural language pairs), enabling it to support a wide range of software engineering tasks, including code summarization, generation, translation, and refinement. By combining unified text-to-text modeling with enhanced pre-training strategies, CodeT5+ provides strong, general-purpose representations for code-related tasks.

\textit{ModernBERT.} ModernBERT~\cite{modernbert} is a re-engineered BERT-style encoder that revisits and modernizes the original BERT architecture using contemporary training practices and optimizations. ModernBERT focuses on improving efficiency and performance through changes such as updated normalization strategies, improved attention implementations, longer context support, and large-scale, high-quality pretraining. The result is a strong, general-purpose encoder that retains the simplicity and interpretability of BERT while achieving substantially better performance and efficiency on modern hardware, making it a competitive backbone for representation learning tasks.
}

\textit{Latent Semantic Indexing (LSI).} LSI~\cite{Deerwester1990} was the first language model to be broadly used for program comprehension~\cite{Maletic1999,Maletic2000}. LSI is used in architectural recovery methods such as~\cite{Maletic1999,Maletic2000,Maletic2001,Marcus2004,Kuhn2005,Kuhn2007}, extract class refactoring techniques such as~\cite{Lucia2008,Bavota2010,Bavota2011,Bavota2013}, coupling metrics including CoCC~\cite{Poshyvanyk2008} and CCBO~\cite{Ujhazi2010}, and cohesion metrics such as C3~\cite{Marcus2005}, LSCM~\cite{Marcus2005} and CLCOM5~\cite{Ujhazi2010}. LSI takes as input a set of documents, each processed into an unordered multiset of terms (i.e., a bag of words). It builds a term-document matrix, typically weighted by TF or TF-IDF, then applies singular value decomposition~\cite{Horn2017} and truncates the result to produce a low-rank approximation. This yields ``term vectors'' and ``document vectors,'' where smaller angles between vectors indicate greater similarity in their (approximate) distributions. As a result, similarity between program entities in an LSI model reflects the extent of their shared vocabulary as projected into a latent semantic space.

\textit{Latent Dirichlet Allocation (LDA).} LDA~\cite{Blei2003} was the next language model to be adopted in program comprehension research~\cite{Baldi2008}. LDA is used in architectural recovery methods such as~\cite{Baldi2008,Garcia2011,Sajnani2012,Shahbazian2018,Zhang2023}, extract class approaches such as~\cite{Akash2019}, and cohesion metrics like MWE~\cite{Liu2009} and NT~\cite{Chen2017}. Like LSI, LDA operates on the term-document matrix, but it treats documents as mixtures of latent topics, where each topic is a distribution over terms. Unlike LSI's algebraic approach, LDA explicitly models the generative process by which documents are composed of topics, making it more suitable for discovering coherent topics in natural language text~\cite{Stevens2012}. Despite explicitly accounting for topics, document vectors in LDA are still ultimately driven by relative term frequencies, as these frequencies determine the posterior estimates of topic proportions.

\textit{Doc2Vec.} Doc2Vec~\cite{Le2014} is not as widely used as LSI or LDA for program comprehension, but it is used in the cohesion metrics COOC and LCOSM~\cite{Miholca2021}. Doc2Vec is a two-layer neural network model trained to predict terms in context using a ``fill in the blank'' training objective. Unlike LSI or LDA, which rely on an unordered bag-of-words representation, Doc2Vec is explicitly trained to capture the sequential context of terms, allowing it to encode word order into embeddings. Despite this improvement, program entities are still considered similar if the lexical content of their definitions is similar; however this does not mean that they address the same concern.

These models represent either the state-of-the-art or those widely used in software engineering tasks. \revision{Other approaches to code embedding include pre-Transformer neural models~\cite{white2016deep,mou2016convolutional,xu2017neural,allamanis2018learning,alon2019code2vec}, which have largely been superseded by Transformer-based models for representation learning tasks. We also excluded decoder-only LLMs such as LLaMA~\cite{CodeLlama} and GPT~\cite{chen2021evaluatinglargelanguagemodels}, as these are aimed at text generation rather than producing embeddings for similarity-based tasks.}

\section{ConcernBERT}\label{sec:concernbert}
In this section, we present ConcernBERT, a model designed to map source code entities into a high-dimensional embedding space that reflects their underlying concerns. We refer to this space as \emph{''concern space''}. We describe the dataset used for training, the model architecture, the training procedure, and relevant implementation details.

\subsection{Dataset}\label{sec:dataset}
ConcernBERT is trained on source code from open source repositories. We obtained the initial list of repositories from The Stack~\cite{Kocetkov2023}, an extensive dataset containing permissively licensed source files from GitHub. This dataset includes over 10 million Java files from 1,016,315 unique GitHub repositories. After consulting the GitHub API, this list was reduced to 713,644 repositories by excluding those that no longer exist or were forks of other repositories. To ensure that we selected repositories that are actively maintained, \revision{we excluded any repository that failed to meet all of the following thresholds: at least five stars, at least five forks, at least five open issues, at least fifty commits, and at least fifty source files.} Additionally, repositories larger than two gigabytes were excluded due to disk space limitations. To focus on real software projects, repositories with names containing keywords such as "example," "exercise," or~"tutorial" were also excluded. \footnote{\revision{The full list of keywords can be found in the replication package.} } Finally, due to tooling limitations, repositories using an encoding other than UTF-8 were excluded. 

After applying these filters, the full list contained \totalrepos repositories, which was partitioned into three sets: \emph{training}, \emph{testing}, and \emph{validation}, containing 5,414 (80\%), 1,082 (16\%), and 270 (4\%) repositories respectively. ConcernBERT was trained exclusively on files from the training set, and we used the validation set to assess model generalization during training. All evaluation is done using the testing set.

Finally, from each repository, we extract a CSV of \emph{essential} source files, excluding any unit tests, integration tests, generated code, or example code. These are identified by matching the path of the source file against a list of keywords such as "test," "example," "gen," etc. The remainder of this paper deals exclusively with essential source files. After removing non-essential files, the train set contains 1,978,071 files across 5,391 repositories; the testing set contains 349,784 files across 1,073 repositories; and the validation set contains 102,277 files across 270 repositories.\footnote{These repository counts are slightly lower than those reported above because a few repositories happened to contain entirely non-essential files.}

\revision{
Intuitively, not all Java classes are small and cohesive. We therefore initially filtered out files that are both large and unstable, defined as having lines of code and change frequency above the 80th percentile. We then conducted experiments (similar to those in Section~\ref{sec:evaluation}) on ConcernBERT variants trained on both the filtered and unfiltered datasets, but observed no measurable difference in the results. In our corpus, the file size at the 80th percentile is approximately 100 LOC, indicating that the vast majority of Java files are relatively small. Since smaller files tend to exhibit higher cohesion, the training data predominantly consists of cohesive classes even without explicit filtering. Consequently, we use the unfiltered dataset in this study.
}

\subsection{Model}
ConcernBERT is trained to map the source code of program entities into a high-dimensional vector space, where entities likely to address similar concerns are positioned close together, while those addressing unrelated concerns are placed further apart. The key innovation of ConcernBERT is that it learns this space from class membership: members of the same Java class, interface, or enum are treated as positive examples (similar concerns), while members of different structures are treated as negative examples (dissimilar concerns). This training strategy enables ConcernBERT to infer concern-level similarity between program entities without relying solely on token-level features.

Existing models typically learn vector representations of program entities directly from their source code. As a result, entities are often judged similar if their textual representations are similar, which does not necessarily reflect their true concerns. For example, \texttt{push()} and \texttt{pop()} methods may share few tokens but are semantically related as stack operations. ConcernBERT uses membership in the same structure to determine similarity during training. This allows the model to capture concern-level relationships beyond surface-level token features, an essential capability for understanding the responsibilities of program entities.

\subsubsection{Architecture}
ConcernBERT builds on the architecture of its base model, CodeBERT~\cite{Feng2020}, by adding a final mean pooling layer to aggregate token embeddings into a single vector representation for each input entity, such as a method or attribute. This modification follows the rationale of SentenceBERT~\cite{Reimers2019}, which adds a mean pooling layer to the original BERT model~\cite{Devlin2019}. In SentenceBERT, token embeddings are aggregated into a single vector representing a sentence; similarly, ConcernBERT aggregates token embeddings into a single vector representing a program entity. In both models, similarity is learned from context: in SentenceBERT, sentences from the same paragraph are considered similar, while in ConcernBERT, entities within the same structure (such as a class, interface, or enum) are assumed to address similar concerns.

\subsubsection{Training}
ConcernBERT is trained using mini-batch stochastic gradient descent. Each mini-batch consists of a random sample of Java classes, interfaces, or enums from the same repository. During training, the model uses triplet loss~\cite{Schroff2015, Hermans2017} to learn the relationships between program entities, operating on triplets of examples: an anchor, a positive, and a negative. In ConcernBERT, the anchor represents a program entity, the positive is another entity from the same structure (indicating similar concerns), and the negative is an entity from a different structure (indicating unrelated concerns). During training, each entity is matched up with the hardest positive from the current batch (the entity from the same structure that is currently the furthest away) and the hardest negative from the current batch (the entity from a different structure that is currently the closest). The triplet loss function ensures that the Euclidean distance between the anchor and the positive entity is minimized, while the Euclidean distance between the anchor and the negative entity is maximized. This encourages ConcernBERT to embed entities with similar concerns close together in the learned vector space, while separating those with dissimilar concerns.

\subsubsection{Implementation}
ConcernBERT was trained over a period of three days on a single NVIDIA A40 GPU. Its initial weights come from \texttt{CodeBERTa-small-v1}~\cite{Husain2019,huggingface_codeberta}, leading to an output dimensionality of 768. Each mini-batch was created by sampling Java structures from the 1,978,071 essential files in the training split, without replacement. Each mini-batch consisted of 166 entities, which was the maximum size that could fit into the available video memory. On average, each batch contained 18.92 Java structures, with each structure containing approximately 8.77 members. ConcernBERT was trained for three epochs where each epoch contained 15,007,396 entities across 90,406 mini-batches. ConcernBERT was implemented using the SentenceTransformers library~\cite{Reimers2019} with minimal hyperparameter tuning.
\begin{figure*}[!t]
\centering
\begin{minipage}{\textwidth}
\centering
\includegraphics{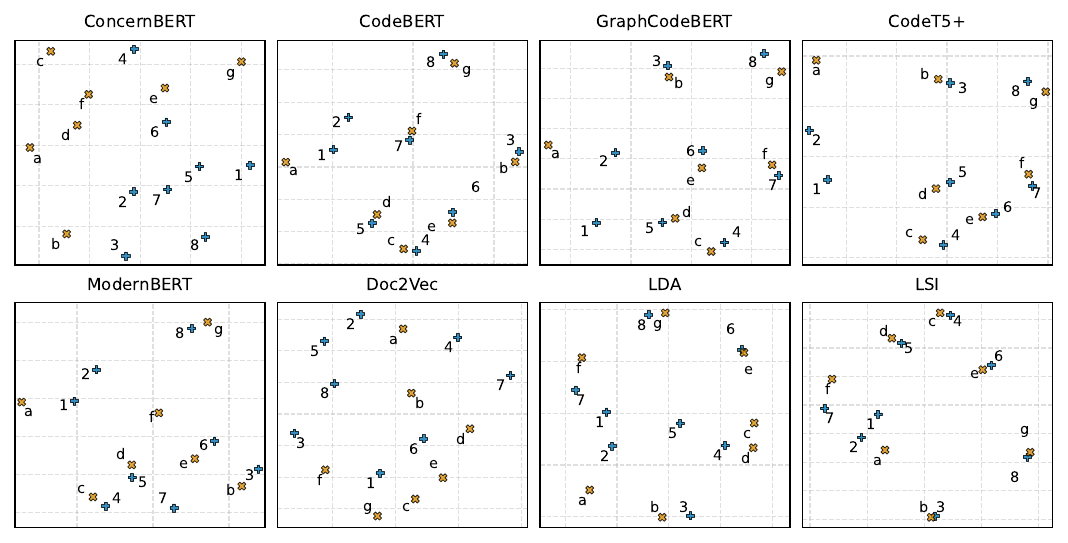}
\end{minipage}

\vspace{1em}

\begin{minipage}{\textwidth}
\centering
\footnotesize
\begin{tabular}{c|l}
\multicolumn{2}{l}{\texttt{src/edu/stanford/nlp/sempre/JoinFormula.java}} \\
\midrule
    1 & \texttt{public JoinFormula(String relation, Formula child) \{ this(Formulas.newNameFormula(relation...} \\
    2 & \texttt{public JoinFormula(Formula relation, Formula child) \{ this.relation = relation; this.child...} \\
    3 & \texttt{public LispTree toLispTree() \{ LispTree tree = LispTree.proto.newList(); tree.addChild(rel...} \\
    4 & \texttt{@Override public void forEach(Function<Formula, Boolean> func) \{ if (!func.apply(this)) \{ ...} \\
    5 & \texttt{@Override public Formula map(Function<Formula, Formula> func) \{ Formula result = func.appl...} \\
    6 & \texttt{@Override public List<Formula> mapToList(Function<Formula, List<Formula>> func, boolean al...} \\
    7 & \texttt{@SuppressWarnings(\{"equalshashcode"\}) @Override public boolean equals(Object thatObj) \{ if...} \\
    \rowcolor{LightYellow}
    8 & \texttt{public int computeHashCode() \{ int hash = 0x7ed55d16; hash = hash * 0xd3a2646c + relation....} \\
\end{tabular}

\vspace{1em}

\begin{tabular}{c|l}
\multicolumn{2}{l}{\texttt{src/edu/stanford/nlp/sempre/NotFormula.java}} \\
\midrule
    a & \texttt{public NotFormula(Formula child) \{ this.child = child; \}} \\
    b & \texttt{public LispTree toLispTree() \{ LispTree tree = LispTree.proto.newList(); tree.addChild("no...} \\
    c & \texttt{@Override public void forEach(Function<Formula, Boolean> func) \{ if (!func.apply(this)) ch...} \\
    d & \texttt{@Override public Formula map(Function<Formula, Formula> func) \{ Formula result = func.appl...} \\
    e & \texttt{@Override public List<Formula> mapToList(Function<Formula, List<Formula>> func, boolean al...} \\
    f & \texttt{@SuppressWarnings(\{"equalshashcode"\}) @Override public boolean equals(Object thatObj) \{ if...} \\
    \rowcolor{LightYellow}
    g & \texttt{public int computeHashCode() \{ int hash = 0x7ed55d16; hash = hash * 0xd3a2646c + child.has...} \\
\end{tabular}
\end{minipage}

\caption{A pair of classes embedded with ConcernBERT and seven baseline models}
\label{fig:embeddings}
\end{figure*}

\section{Example}
In this section, we illustrate the novelty of ConcernBERT using a running example. Figure~\ref{fig:embeddings} presents two classes, \texttt{JoinFormula} and \texttt{NotFormula}, which contain 8 and 7 methods, respectively. Although these two classes have distinct responsibilities, many of their tokens are shared, such as \texttt{Formula}, \texttt{List}, \texttt{map}, and \texttt{child}. For inexperienced developers or those under deadline pressures, these responsibilities could be mistakenly combined into a single, overly complex class containing various formula-related concerns.

To simulate this scenario, we mix all 15 methods into a synthetic class and evaluate how well ConcernBERT distinguishes between their original responsibilities, in comparison to the baseline models. We generate embeddings using all eight models, project them into two dimensions using Multidimensional Scaling (MDS)~\cite{Kruskal1964}, and visualize them as scatterplots in Figure~\ref{fig:embeddings}. In each plot, methods from \texttt{JoinFormula} are shown as blue plus signs indexed with numbers, while methods from \texttt{NotFormula} are shown as orange crosses indexed with letters.

The ConcernBERT scatterplot clearly separates the two groups of methods based on their original class membership. For example, methods ``8'' and ``g'' both implement \texttt{computeHashCode} using nearly identical tokens in the same order. In the baseline models, these methods are typically embedded close together: CodeBERT, GraphCodeBERT, ModernBERT, and LDA all place ``8'' and ``g'' in nearly the same position. This pattern reflects token-level similarity but fails to capture the distinct class responsibilities. In contrast, ConcernBERT correctly positions ``8'' closer to methods from its original class (such as ``1,'' ``2,'' and ``5'') than to ``g.'' While LSI separates ``8'' and ``g,'' it does not achieve clear separation between the two classes overall.

This demonstrates the advantage that ConcernBERT gains from being trained on class membership: all methods in \texttt{JoinFormula} refer to both attributes \texttt{relation} and \texttt{child}, whereas methods in \texttt{NotFormula} only use the \texttt{child} attribute. This distinction is uniquely encoded in ConcernBERT, enabling it to differentiate responsibilities even when token-level similarity is high. This aligns with the intuition behind the original LCOM~\cite{Chidamber1991} metric, which defines cohesion in terms of shared attribute references. Training on class context allows ConcernBERT to capture such semantic cohesion.
\section{Evaluation}~\label{sec:evaluation}
We now evaluate the performance of ConcernBERT, addressing the following research question:

\textbf{RQ:} \emph{Does ConcernBERT outperform existing models in distinguishing distinct responsibilities among program entities, and how do clustering algorithm and distance metric choices affect this performance?}

This question evaluates ConcernBERT's effectiveness in recovering meaningful responsibilities from program entities, an essential step for downstream tasks like architectural recovery, extract class refactoring, and coupling and cohesion metrics. We further investigate whether its improvements stem from its specialized training process.

We compare ConcernBERT against seven reference models. The first three, LSI, LDA, and Doc2Vec, are commonly used in tasks such as extract class refactoring, architecture recovery, and software quality analysis. The remaining four are Transformer-based models: CodeBERT, which serves as the architectural foundation for ConcernBERT; GraphCodeBERT, which incorporates data flow information; CodeT5+, an encoder-decoder model; and ModernBERT, a recent encoder model with architectural improvements. These models were briefly introduced in Section~\ref{sec:related-work}. In this section, we describe the evaluation process and present the results.

\begin{figure*}[htbp]
\centering
\includegraphics[width=0.8\textwidth]{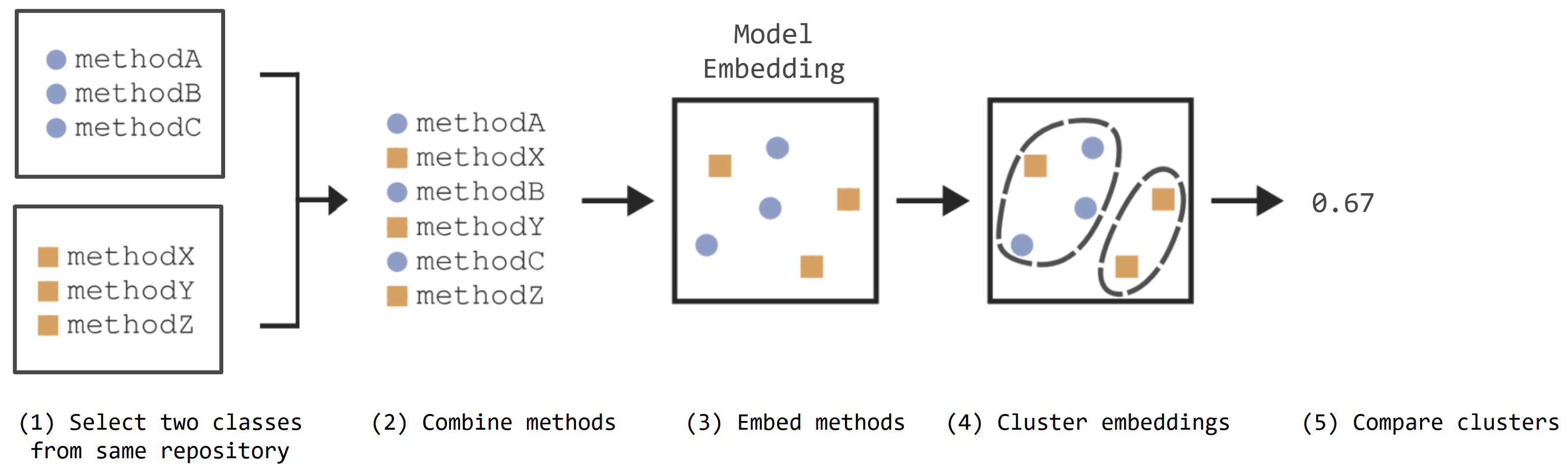}
\caption{Illustration of the Class-Membership Recovery Test when $k=2$}
\label{fig:membership-recovery}
\end{figure*}

\subsection{Evaluation Process}
We evaluate ConcernBERT and the reference models using repeated trials of the \emph{Class-Membership Recovery Test}, illustrated in Figure~\ref{fig:membership-recovery}. In each trial, we randomly select $k$ classes from the same repository and merge their methods into a single unlabeled group. Each model then encodes these methods into its respective embedding space, and the embeddings are clustered into $k$ clusters. The resulting clusters are compared against the original class memberships, which serve as the ground truth. A model that more accurately recovers the original class structure is considered better at distinguishing program concerns. This test closely simulates real-world tasks such as extract class refactoring, where the goal is to identify cohesive subsets of methods that can be separated into distinct classes.

We evaluate performance across four group sizes~(\mbox{$k \in \{2, 10, 50, 100\}$}) to assess how models handle varying levels of complexity. Smaller values of $k$ represent simpler scenarios with fewer responsibilities to distinguish, while larger values represent more challenging cases where many interleaved concerns must be disentangled. Since clustering results can vary depending on the algorithm and distance metric used, we evaluate each model across twelve configurations (four clustering algorithms combined with three distance metrics), allowing each model to be assessed under its most favorable configuration. Next, we elaborate on each step of this process.

\subsubsection{Train Reference Models}
While ConcernBERT, CodeBERT, GraphCodeBERT, CodeT5+, and ModernBERT can be used directly, LSI, LDA, and Doc2Vec require training on a specific corpus to generate embeddings. Following prior work (e.g.,~\cite{Maletic1999,Marcus2004,Liu2009,Miholca2021}), we construct a separate corpus for each repository, consisting of the text from all methods within that project. Methods are preprocessed to retain only comments, identifiers, and string literals; English stop words are removed, and both stemming and lemmatization are applied. LSI, LDA, and Doc2Vec are then trained on these corpora using the Gensim~\cite{Rehurek2010} library, a widely used Python toolkit for topic modeling and document similarity. We use Gensim's default hyperparameters, except for the embedding dimension, which is set to 768 to match the output size of the Transformer-based models. This ensures that all models produce embeddings of the same dimensionality, enabling a fair comparison.

\subsubsection{Create Tests}
Next, we construct test instances by merging the methods of randomly selected classes. As described in Section~\ref{sec:dataset}, we split the 6,766 repositories into training, validation, and testing sets, containing 5,391, 270, and 1,073 repositories, respectively. For evaluation, we select classes exclusively from the testing set to ensure that no evaluated class was seen during ConcernBERT's training. Classes are merged only if they come from the same repository. We include only essential classes: those that are not tests, examples, or generated code.

\revision{To test model performance across a range of scenarios, we vary the number of classes merged per test instance, using $k \in \{2, 10, 50, 100\}$. For each value of $k$, we repeatedly sample $k$ classes from the same repository (without replacement) and merge their methods into a single unlabeled group. This yields 109,489 test instances in total: 88,027 for $k=2$, 17,173 for $k=10$, 3,006 for $k=50$, and 1,283 for $k=100$, spanning 176,560 unique classes.} As discussed in Section~\ref{sec:threats}, grouping distinct classes represents an ideal case. While real-world classes may mix interrelated responsibilities, our rationale is that models failing on clear-cut cases are unlikely to succeed on more ambiguous ones.

\subsubsection{Embed Methods}
Next, the methods from all test instances are processed by the models, with each model producing a 768-dimensional embedding for each method. Our evaluation includes a total of 1,195,679 unique methods.

\subsubsection{Cluster Methods}
Next, given the method embeddings produced by each model for each test, we cluster them into $k$ clusters (matching the number of classes in each group) to assess how well the resulting clusters recover the original class memberships. Since results may vary depending on the clustering algorithm and distance metric, our evaluation includes four clustering methods and three distance measures, all commonly used in prior work, resulting in 12 total configurations. This setup allows us to examine how sensitive each model's performance is to different clustering strategies and distance functions. We briefly introduce these methods below.

We use K-medoids clustering\cite{Schubert2021} and three variants of Hierarchical Agglomerative Clustering~(HAC)\cite{Hastie2009}. K-medoids is representative of partitional clustering and has found success in prior work~\cite{Corazza2011}, while HAC, representative of hierarchical clustering, is the most widely used across downstream tasks~\cite{Kuhn2005, Kuhn2007, Maqbool2007, Corazza2011, Puchala2022, Akash2019, Fokaefs2012}. K-medoids partitions the data into a fixed number of clusters by minimizing the sum of distances to the cluster medoids. Unlike the original K-means algorithm, which uses squared distances to cluster means, K-medoids supports arbitrary distance metrics. HAC constructs a hierarchy by iteratively merging the two closest clusters based on a linkage criterion. We consider three standard variants: single linkage (minimum distance between points), average linkage (mean pairwise distance), and complete linkage (maximum distance between points), each capturing a different notion of inter-cluster similarity. 

Each clustering method is combined with one of three distance metrics: cosine distance, Euclidean distance, and Jensen-Shannon distance (JSD)~\cite{Lin1991}. Cosine distance measures the angle between vectors and is commonly used for comparing high-dimensional embeddings. Euclidean distance measures straight-line distance in a vector space and aligns with ConcernBERT's training objective, which minimizes pairwise Euclidean distances among related entities. JSD, a symmetric and smoothed variant of Kullback-Leibler divergence, is well-suited for comparing probability distributions and is therefore appropriate for models like LDA, which output such distributions.

\subsubsection{Measure Performance}
Finally, to measure the performance of each model, we assess the agreement between the clusters it produces and the original class membership. In this step, we use Adjusted Mutual Information~(AMI)~\cite{Nguyen2010}, which quantifies the similarity between two clusterings while correcting for chance agreement. We chose AMI because it is a state-of-the-art clustering validation metric, and combines normalization, chance correction, and robustness to cluster size and count, making it more reliable and statistically sound. MoJoFM~\cite{Wen2004} is also widely used in architectural recovery and extract class refactoring research; it measures the minimum number of move and join operations needed to transform one clustering into another. It is not normalized and is sensitive to the number and size of clusters. We also explored other metrics, including Normalized Mutual Information~(NMI)~\cite{Strehl2002}, which measures mutual dependence but lacks chance correction; Rand Index~(RI)~\cite{Rand1971}, which computes pairwise agreement without accounting for randomness; and Adjusted Rand Index (ARI)~\cite{Hubert1985}, which adjusts for chance but is sensitive to the number of clusters. 

All results are available in our replication package~\cite{Lefever2025a}. We observe that the specific choice of validation metric has minimal impact, as all metrics produce consistent rankings across models.

\begin{table*}
  \centering
  \setlength{\extrarowheight}{0pt}
  \addtolength{\extrarowheight}{\aboverulesep}
  \setlength{\aboverulesep}{0pt}
  \setlength{\belowrulesep}{0pt}
  \caption{Class-Membership Recovery Results}
  \label{tbl:cmr-results}
  \begin{tabular}{%
      l || 
      *{3}{r} | 
      *{3}{r} | 
      *{3}{r} | 
      *{3}{r}   
    }
    \toprule
    & \multicolumn{3}{c|}{K-Medoids} & \multicolumn{3}{c|}{HAC (Single)} & \multicolumn{3}{c|}{HAC (Average)} & \multicolumn{3}{c}{HAC (Complete)} \\
    Model & Cos & Euc & JSD & Cos & Euc & JSD & Cos & Euc & JSD & Cos & Euc & JSD \\
    \midrule
    \multicolumn{13}{c}{\textbf{$k=2$}} \\
    \midrule
    ConcernBERT    & \cellcolor{LightYellow}\underline{0.728} & \cellcolor{LightYellow}0.709 & \cellcolor{LightYellow}0.680 & \cellcolor{LightYellow}0.583 & \cellcolor{LightYellow}0.584 & \cellcolor{LightYellow}0.555 & \cellcolor{LightYellow}0.693 & \cellcolor{LightYellow}0.692 & \cellcolor{LightYellow}0.657 & \cellcolor{LightYellow}0.720 & \cellcolor{LightYellow}0.720 & \cellcolor{LightYellow}0.676 \\
    CodeBERT       & \underline{0.338} & 0.284 & 0.319 & 0.245 & 0.179 & 0.215 & 0.318 & 0.228 & 0.303 & 0.320 & 0.239 & 0.325 \\
    GraphCodeBERT  & \underline{0.284} & 0.281 & 0.230 & 0.201 & 0.200 & 0.154 & 0.239 & 0.238 & 0.180 & 0.241 & 0.240 & 0.184 \\
    CodeT5+        & 0.226 & \underline{0.227} & 0.199 & 0.150 & 0.159 & 0.110 & 0.179 & 0.184 & 0.140 & 0.185 & 0.184 & 0.157 \\
    ModernBERT     & \underline{0.246} & 0.245 & 0.228 & 0.172 & 0.172 & 0.159 & 0.197 & 0.198 & 0.177 & 0.196 & 0.196 & 0.176 \\
    Doc2Vec        & \underline{0.190} & 0.147 & 0.183 & 0.114 & 0.087 & 0.101 & 0.161 & 0.112 & 0.146 & 0.184 & 0.150 & 0.180 \\
    LDA            & 0.472 & 0.302 & \underline{0.475} & 0.369 & 0.139 & 0.402 & 0.453 & 0.173 & 0.476 & 0.435 & 0.206 & 0.429 \\
    LSI            & \underline{0.558} & 0.152 & 0.352 & 0.440 & 0.073 & 0.148 & 0.552 & 0.094 & 0.201 & 0.510 & 0.125 & 0.266 \\
    \midrule
    \multicolumn{13}{c}{\textbf{$k=10$}} \\
    \midrule
    ConcernBERT    & \cellcolor{LightYellow}0.659 & \cellcolor{LightYellow}0.644 & \cellcolor{LightYellow}0.623 & \cellcolor{LightYellow}0.441 & \cellcolor{LightYellow}0.442 & \cellcolor{LightYellow}0.423 & \cellcolor{LightYellow}0.625 & \cellcolor{LightYellow}0.624 & \cellcolor{LightYellow}0.599 & \cellcolor{LightYellow}\underline{0.669} & \cellcolor{LightYellow}0.669 & \cellcolor{LightYellow}0.643 \\
    CodeBERT       & \underline{0.365} & 0.339 & 0.342 & 0.233 & 0.173 & 0.187 & 0.355 & 0.284 & 0.331 & 0.360 & 0.320 & 0.350 \\
    GraphCodeBERT  & \underline{0.330} & 0.325 & 0.298 & 0.203 & 0.204 & 0.181 & 0.306 & 0.307 & 0.276 & 0.316 & 0.315 & 0.278 \\
    CodeT5+        & \underline{0.315} & 0.310 & 0.280 & 0.173 & 0.187 & 0.109 & 0.267 & 0.276 & 0.196 & 0.287 & 0.292 & 0.239 \\
    ModernBERT     & \underline{0.325} & 0.317 & 0.312 & 0.184 & 0.183 & 0.174 & 0.301 & 0.301 & 0.284 & 0.309 & 0.309 & 0.297 \\
    Doc2Vec        & \underline{0.227} & 0.154 & 0.218 & 0.080 & 0.085 & 0.069 & 0.173 & 0.109 & 0.153 & 0.230 & 0.154 & 0.224 \\
    LDA            & 0.451 & 0.334 & \underline{0.461} & 0.299 & 0.121 & 0.340 & 0.469 & 0.184 & 0.502 & 0.416 & 0.241 & 0.438 \\
    LSI            & \underline{0.498} & 0.193 & 0.371 & 0.334 & 0.063 & 0.102 & 0.504 & 0.084 & 0.171 & 0.476 & 0.139 & 0.284 \\
    \midrule
    \multicolumn{13}{c}{\textbf{$k=50$}} \\
    \midrule
    ConcernBERT    & \cellcolor{LightYellow}0.571 & \cellcolor{LightYellow}0.556 & \cellcolor{LightYellow}0.538 & \cellcolor{LightYellow}0.298 & \cellcolor{LightYellow}0.299 & \cellcolor{LightYellow}0.287 & \cellcolor{LightYellow}0.557 & \cellcolor{LightYellow}0.556 & \cellcolor{LightYellow}0.537 & \cellcolor{LightYellow}\underline{0.599} & \cellcolor{LightYellow}0.600 & \cellcolor{LightYellow}0.579 \\
    CodeBERT       & \underline{0.307} & 0.288 & 0.286 & 0.149 & 0.115 & 0.115 & 0.315 & 0.266 & 0.291 & 0.319 & 0.296 & 0.308 \\
    GraphCodeBERT  & \underline{0.284} & 0.277 & 0.260 & 0.138 & 0.139 & 0.126 & 0.269 & 0.268 & 0.262 & 0.284 & 0.283 & 0.266 \\
    CodeT5+        & \underline{0.281} & 0.273 & 0.250 & 0.123 & 0.135 & 0.074 & 0.256 & 0.264 & 0.191 & 0.278 & 0.282 & 0.233 \\
    ModernBERT     & \underline{0.283} & 0.273 & 0.271 & 0.119 & 0.119 & 0.115 & 0.287 & 0.286 & 0.279 & 0.289 & 0.288 & 0.285 \\
    Doc2Vec        & \underline{0.187} & 0.117 & 0.175 & 0.047 & 0.056 & 0.038 & 0.146 & 0.073 & 0.130 & 0.190 & 0.114 & 0.185 \\
    LDA            & 0.350 & 0.272 & 0.368 & 0.200 & 0.096 & 0.215 & 0.369 & 0.223 & \underline{0.408} & 0.346 & 0.258 & 0.379 \\
    LSI            & \underline{0.390} & 0.199 & 0.319 & 0.195 & 0.043 & 0.066 & 0.404 & 0.067 & 0.152 & 0.390 & 0.132 & 0.272 \\
    \midrule
    \multicolumn{13}{c}{\textbf{$k=100$}} \\
    \midrule
    ConcernBERT    & \cellcolor{LightYellow}0.532 & \cellcolor{LightYellow}0.518 & \cellcolor{LightYellow}0.501 & \cellcolor{LightYellow}0.254 & \cellcolor{LightYellow}0.255 & \cellcolor{LightYellow}0.247 & \cellcolor{LightYellow}0.531 & \cellcolor{LightYellow}0.530 & \cellcolor{LightYellow}0.513 & \cellcolor{LightYellow}\underline{0.565} & \cellcolor{LightYellow}0.565 & \cellcolor{LightYellow}0.547 \\
    CodeBERT       & 0.275 & 0.259 & 0.255 & 0.121 & 0.095 & 0.096 & \underline{0.292} & 0.250 & 0.271 & 0.291 & 0.273 & 0.281 \\
    GraphCodeBERT  & \underline{0.256} & 0.249 & 0.234 & 0.116 & 0.118 & 0.107 & 0.249 & 0.248 & 0.246 & 0.262 & 0.261 & 0.247 \\
    CodeT5+        & 0.256 & 0.247 & 0.227 & 0.104 & 0.116 & 0.063 & 0.244 & 0.252 & 0.183 & 0.261 & \underline{0.263} & 0.220 \\
    ModernBERT     & \underline{0.256} & 0.246 & 0.245 & 0.101 & 0.100 & 0.097 & 0.269 & 0.268 & 0.265 & 0.268 & 0.267 & 0.266 \\
    Doc2Vec        & \underline{0.167} & 0.099 & 0.155 & 0.038 & 0.045 & 0.030 & 0.133 & 0.058 & 0.119 & 0.171 & 0.099 & 0.166 \\
    LDA            & 0.304 & 0.246 & 0.324 & 0.170 & 0.088 & 0.168 & 0.321 & 0.248 & \underline{0.364} & 0.311 & 0.264 & 0.345 \\
    LSI            & \underline{0.343} & 0.196 & 0.290 & 0.155 & 0.036 & 0.056 & 0.365 & 0.061 & 0.153 & 0.353 & 0.128 & 0.270 \\
    \bottomrule
  \end{tabular}
\end{table*}

\subsection{Results}
We report the results of the Class-Membership Recovery Tests in Table~\ref{tbl:cmr-results}. The table is organized into four sections by group size ($k \in \{2, 10, 50, 100\}$). Within each section, rows correspond to the eight models and columns correspond to the twelve clustering configurations. Each cell reports the mean AMI across all trials for the corresponding model-configuration pair. The best-performing model for each configuration is highlighted in yellow, while each model's best-performing configuration is underlined. A higher AMI indicates better recovery of the original class memberships.

\subsubsection{Statistical Analysis}
To compare embedding models, we pair each model with its best-performing clustering configuration (underlined in Table~\ref{tbl:cmr-results}). \revision{Results are reported in Table~\ref{tbl:cmr-summary}, which presents the mean AMI, 95\% confidence interval (computed using standard error), and Cliff's delta~\cite{Cliff1993} effect size relative to ConcernBERT. Cliff's delta is a non-parametric measure ranging from $-1$ to $+1$, where positive values indicate ConcernBERT outperforms the comparison model. Following standard conventions~\cite{Romano2006}, we interpret $|\delta| < 0.147$ as negligible, $0.147 \leq |\delta| < 0.33$ as small, $0.33 \leq |\delta| < 0.474$ as medium, and $|\delta| \geq 0.474$ as large.}

\revision{ConcernBERT substantially outperforms all reference models across all group sizes, with effect sizes that are large for most comparisons and grow larger as task difficulty increases. For $k=2$, ConcernBERT achieves a mean AMI of 0.728, outperforming LSI (the next best model) by 0.170 points ($\delta = 0.22$). The improvements over the Transformer models are larger: CodeBERT (0.390 points, $\delta = 0.49$), GraphCodeBERT (0.444 points, $\delta = 0.55$), ModernBERT (0.482 points, $\delta = 0.60$), and CodeT5+ (0.501 points, $\delta = 0.62$). ConcernBERT's advantage grows as task difficulty increases: for $k=100$, it achieves a mean of 0.565, outperforming LSI by 0.200 points ($\delta = 0.82$) and LDA by 0.201 points ($\delta = 0.81$). All pairwise comparisons are statistically significant (Wilcoxon signed-rank test, $p < 0.01$).}

\revision{Notably, effect sizes increase with $k$. At $k=2$, the task is relatively easy and all models perform reasonably well, compressing the differences between them. As $k$ increases, the task becomes more challenging, and ConcernBERT's advantage becomes more pronounced: for example, the effect size relative to CodeBERT grows from $\delta = 0.49$ at $k=2$ to $\delta = 0.91$ at $k=100$. This pattern indicates that ConcernBERT's superiority is most evident on difficult tasks, where distinguishing many interleaved concerns requires deeper semantic understanding.}

\revision{The comparison between ConcernBERT and CodeBERT is particularly informative, as both models share the same architecture but differ in their training process. At $k=2$, ConcernBERT outperforms CodeBERT by 0.390 points ($\delta = 0.49$), and at $k=100$ by 0.273 points ($\delta = 0.91$). The large effect sizes confirm that ConcernBERT's advantage stems from its membership-based training rather than architectural factors. Notably, ConcernBERT also substantially outperforms the more recent Transformer models (GraphCodeBERT, CodeT5+, and ModernBERT) across all group sizes. These results show that architectural advances alone do not capture concern-level semantics; ConcernBERT's training process is the key to its superior performance.}


\subsubsection{Result Distribution and Trends}
\begin{figure*}
\centering
\includegraphics{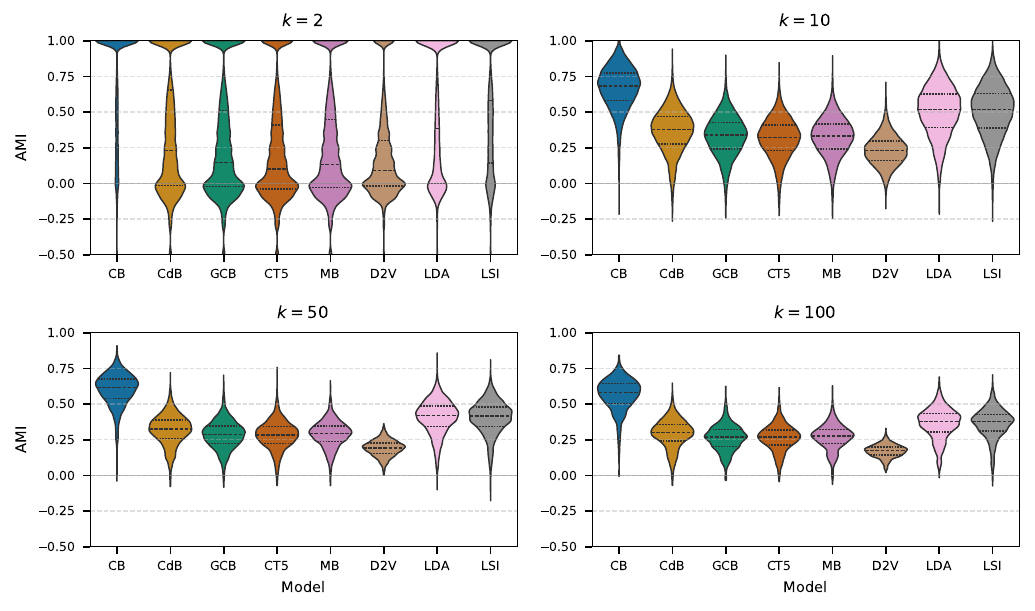}
\caption{Distribution of AMI scores across different models for varying test sizes ($k$). Horizontal lines within violins indicate quartiles. Model abbreviations: CB = ConcernBERT, CdB = CodeBERT, GCB = GraphCodeBERT, CT5 = CodeT5+, MB = ModernBERT, D2V = Doc2Vec.}
\label{fig:recovery_violin_plots}
\end{figure*}

To assess the consistency and reliability of each model, we use violin plots (Figure~\ref{fig:recovery_violin_plots}) to visualize the distribution of AMI scores across test cases for all four group sizes. While average scores summarize overall performance, these plots reveal how performance varies: whether a model performs well consistently or only in specific cases. In each violin plot, the width at any vertical position reflects the density of scores: wider regions indicate a higher concentration of values. The horizontal lines within each violin indicate quartiles, allowing us to evaluate not just central tendency, but also the spread and shape of each model's performance distribution.

\begin{table*}
  \centering
  \setlength{\extrarowheight}{0pt}
  \addtolength{\extrarowheight}{\aboverulesep}
  \setlength{\aboverulesep}{0pt}
  \setlength{\belowrulesep}{0pt}
  \caption{Summary of Model Performance}
  \label{tbl:cmr-summary}
  \begin{tabular}{%
      l || 
      *{3}{r} | 
      *{3}{r} | 
      *{3}{r} | 
      *{3}{r}   
    }
    \toprule
    & \multicolumn{3}{c|}{$k=2$} & \multicolumn{3}{c|}{$k=10$} & \multicolumn{3}{c|}{$k=50$} & \multicolumn{3}{c}{$k=100$} \\
    Model & Mean & 95\% CI & $\delta$ & Mean & 95\% CI & $\delta$ & Mean & 95\% CI & $\delta$ & Mean & 95\% CI & $\delta$ \\
    \midrule
    \cellcolor{LightYellow}ConcernBERT & \cellcolor{LightYellow}0.728 & \cellcolor{LightYellow}[.726, .731] & \cellcolor{LightYellow}--- & \cellcolor{LightYellow}0.669 & \cellcolor{LightYellow}[.667, .671] & \cellcolor{LightYellow}--- & \cellcolor{LightYellow}0.600 & \cellcolor{LightYellow}[.595, .604] & \cellcolor{LightYellow}--- & \cellcolor{LightYellow}0.565 & \cellcolor{LightYellow}[.559, .571] & \cellcolor{LightYellow}--- \\
    CodeBERT      & 0.338 & [.335, .341] & 0.49 & 0.365 & [.363, .368] & 0.85 & 0.319 & [.315, .323] & 0.91 & 0.292 & [.287, .298] & 0.91 \\
    GraphCodeBERT & 0.284 & [.281, .286] & 0.55 & 0.330 & [.328, .332] & 0.89 & 0.284 & [.281, .288] & 0.94 & 0.262 & [.257, .267] & 0.94 \\
    CodeT5+       & 0.227 & [.224, .229] & 0.62 & 0.315 & [.313, .317] & 0.91 & 0.282 & [.278, .285] & 0.95 & 0.263 & [.258, .268] & 0.94 \\
    ModernBERT    & 0.246 & [.244, .249] & 0.60 & 0.325 & [.323, .327] & 0.91 & 0.289 & [.286, .292] & 0.94 & 0.269 & [.264, .274] & 0.94 \\
    Doc2Vec       & 0.190 & [.188, .192] & 0.67 & 0.230 & [.229, .232] & 0.97 & 0.190 & [.188, .192] & 0.99 & 0.171 & [.169, .173] & 0.98 \\
    LDA           & 0.476 & [.473, .479] & 0.31 & 0.502 & [.499, .505] & 0.54 & 0.408 & [.403, .412] & 0.78 & 0.364 & [.358, .370] & 0.81 \\
    LSI           & 0.558 & [.555, .560] & 0.22 & 0.504 & [.501, .506] & 0.54 & 0.404 & [.400, .408] & 0.79 & 0.365 & [.359, .370] & 0.82 \\
    \bottomrule
  \end{tabular}
\end{table*}

For the small-group tests ($k=2$), ConcernBERT exhibits a tight, narrow distribution concentrated near 1.0, indicating highly accurate and consistent clustering. In contrast, the baseline models display broader, flatter distributions with lower medians and long tails extending below zero. An AMI score below zero indicates worse-than-random clustering, meaning that the predicted groupings are less aligned with the ground truth than would be expected by chance.

\begin{figure}[t]
\centering
\includegraphics{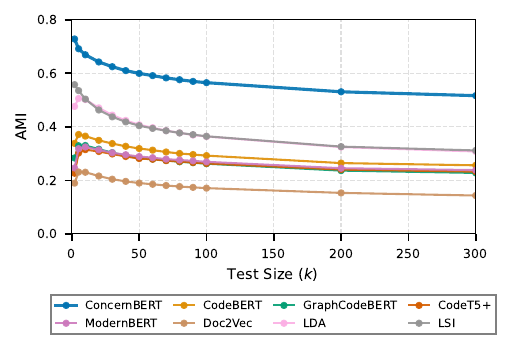}
\caption{Mean AMI scores as a function of test size ($k$)}
\label{fig:ami_vs_k}
\end{figure}

\begin{figure*}[!t]
\centering
\begin{subfigure}[t]{0.32\textwidth}
    \centering
    \includegraphics{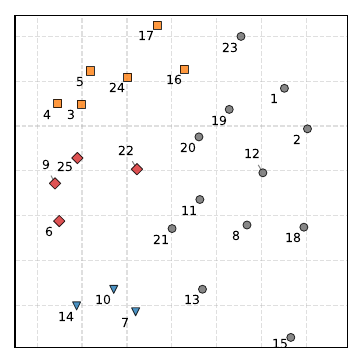}
    \caption{ConcernBERT}
    \label{fig:case-study-concernbert}
\end{subfigure}
\hfill
\begin{subfigure}[t]{0.32\textwidth}
    \centering
    \includegraphics{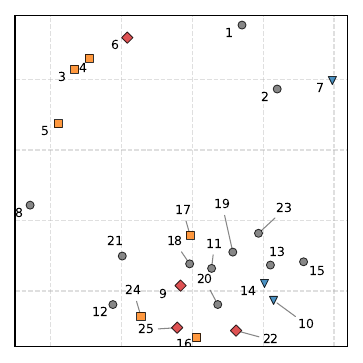}
    \caption{CodeBERT}
    \label{fig:case-study-codebert}
\end{subfigure}
\hfill
\begin{subfigure}[t]{0.32\textwidth}
    \centering
    \includegraphics{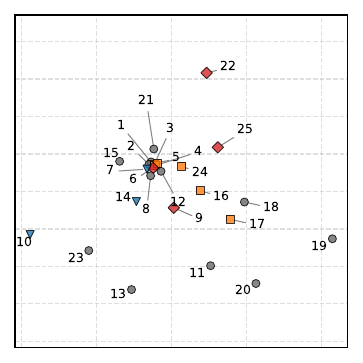}
    \caption{LSI}
    \label{fig:case-study-lsi}
\end{subfigure}

\vspace{1em}

\begin{subfigure}[t]{\textwidth}
    \centering
    \footnotesize
    \begin{tabular}{c|l}
    \multicolumn{2}{l}{\texttt{java/tools/src/main/java/org/apache/tsfile/tools/TsFileTool.java}} \\
    \midrule
    1 & \texttt{private static int THREAD\_COUNT = 8;} \\
    2 & \texttt{private static long CHUNK\_SIZE\_BYTE = 1024 * 1024 * 256;} \\
    3 & \texttt{private static String outputDirectoryStr = "";} \\
    4 & \texttt{private static String inputDirectoryStr = "";} \\
    5 & \texttt{private static String failedDirectoryStr = "failed";} \\
    6 & \texttt{private static String schemaPathStr = "";} \\
    7 & \texttt{private static SchemaParser.Schema schema = null;} \\
    8 & \texttt{private static final Logger LOGGER = LoggerFactory.getLogger(TsFileTool.class);} \\
    9 & \texttt{public static void main(String[] args) \{ if (System.getenv("TSFILE\_HOME") != null) \{ Syste...} \\
    10 & \texttt{private static TableSchema genTableSchema( List<SchemaParser.IDColumns> idColumnList, List...} \\
    11 & \texttt{private static boolean writeTsFile( String sourceFilePath, String fileName, List<String> l...} \\
    12 & \texttt{private static void deleteFile(File tsfile) \{ if (!tsfile.delete()) \{ LOGGER.error(tsfile....} \\
    13 & \texttt{private static Tablet genTablet( TableSchema tableSchema, List<String> lineList, Map<Strin...} \\
    14 & \texttt{public static List<String[]> sortAndParseLines(List<String> data) \{ List<String[]> parsedL...} \\
    15 & \texttt{public static Object getValue(TSDataType dataType, String i, ColumnCategory columnCategory...} \\
    16 & \texttt{private static void processDirectory(File directory, ExecutorService executor) \{ if (direc...} \\
    17 & \texttt{private static void cpFile(String sourceFilePath, String targetDirectoryPath) \{ try \{ Stri...} \\
    18 & \texttt{public static void writeToNewCSV( String headerLine, String fileAbsolutePath, List<String>...} \\
    19 & \texttt{private static void processFile(File inputFile, ExecutorService executor) \{ AtomicInteger ...} \\
    20 & \texttt{private static void submitChunk( String headerLine, List<String> lineList, int fileNumber,...} \\
    21 & \texttt{private static void printHelp(Options options) \{ HelpFormatter formatter = new HelpFormatt...} \\
    22 & \texttt{private static void parseCommandLineParams(String[] args) \{ Options options = new Options(...} \\
    23 & \texttt{private static long parseBlockSize(String blockSizeValue) \{ long size; blockSizeValue = bl...} \\
    24 & \texttt{private static void createDir() \{ File targetDir = new File(outputDirectoryStr); if (!targ...} \\
    25 & \texttt{private static boolean validateParams() \{ if (inputDirectoryStr == null || inputDirectoryS...} \\
    \end{tabular}
    \caption{Source code of each entity in the class}
    \label{fig:case-study-entities}
\end{subfigure}

\caption{Comparison of embedding results for the \texttt{TsFileTool} class}
\label{fig:case-study}
\end{figure*}

As task difficulty increases from $k=2$ to $k=100$, all models show degraded performance, reflected in distributions that shift downward and become more dispersed. However, ConcernBERT maintains the highest and most compact distribution across all group sizes. For $k=100$, ConcernBERT's scores remain clustered around 0.55--0.60, while baseline models show substantially lower performance. Doc2Vec exhibits a particularly tight cluster of low scores around 0.15--0.20, while other baseline models show more variation but still fall considerably short of ConcernBERT. These results demonstrate that although all models degrade under more difficult conditions, ConcernBERT remains the most accurate and consistent across all test scenarios, from simple two-class cases to complex 100-class mixtures.

\revision{Figure~\ref{fig:ami_vs_k} further illustrates how model performance evolves as the number of merged classes increases from $k=2$ to $k=300$. All models exhibit performance degradation as task complexity grows, but ConcernBERT demonstrates the most gradual decline, maintaining a mean AMI above 0.5 even at $k=300$. Notably, the performance gap between ConcernBERT and all baseline models widens as $k$ increases, indicating that ConcernBERT's advantage becomes more pronounced in challenging scenarios requiring the disentanglement of many interleaved concerns. The trend reveals three distinct and stable performance tiers: ConcernBERT at the top, LSI and LDA in the middle, and the remaining models (CodeBERT, GraphCodeBERT, CodeT5+, ModernBERT, and Doc2Vec) clustered at the bottom. This consistent ordering across all test sizes reinforces that ConcernBERT's superiority holds across diverse levels of complexity.}

\revision{
\subsubsection{Runtime Performance}
To assess ConcernBERT's practical feasibility for integration with development tools, we measured embedding times on 50 projects from our test set. On a machine equipped with an NVIDIA A40 GPU and an AMD EPYC 7662 CPU, ConcernBERT embeds methods at approximately 456 methods per second, comparable to CodeBERT (457 methods per second). The larger Transformer models exhibit lower throughput: GraphCodeBERT (209 methods per second), CodeT5+ (203 methods per second), and ModernBERT (151 methods per second). The traditional models, when accounting for per-project training time, achieve variable throughput: LSI (904 methods per second), Doc2Vec (505 methods per second), and LDA (195 methods per second). These timings suggest that ConcernBERT is sufficiently efficient for interactive use cases such as IDE-based refactoring suggestions. For batch processing scenarios such as continuous integration pipelines, the overhead is negligible relative to typical build times, and incremental re-embedding of only modified files can further reduce costs in practice.
}

\revision{
\subsubsection{Case Study}
The promising results of the statistical analysis give us confidence that ConcernBERT can significantly improve the efficiency of tasks that require identifying \emph{cohesive entity groups}, such as god class decomposition and class-level refactoring extraction. To demonstrate the potential and advantages of ConcernBERT in identifying cohesive entity groups, we present a comparative study on decomposing \texttt{TsFileTool}, a complex class from Apache TsFile.\footnote{The full source code is available \href{https://github.com/apache/tsfile/blob/9c2711ca0d81d1b6523b1b5d36de594d48b35d11/java/tools/src/main/java/org/apache/tsfile/tools/TsFileTool.java}{here} on GitHub.} \minor{This case study is intended to illustrate that ConcernBERT can  provide improved refactoring recommendations, rather than serving as a comprehensive evaluation of end-to-end refactoring capability.}

Figure~\ref{fig:case-study} visualizes the 25 entities of \texttt{TsFileTool.java} when embedded using ConcernBERT, CodeBERT, and LSI, with the source code shown in Figure~\ref{fig:case-study-entities}. We compare ConcernBERT against CodeBERT, the best-performing Transformer baseline, and LSI, the best peforming of the traditional models.

Figure~\ref{fig:case-study-concernbert} presents the embeddings generated by ConcernBERT, which identifies several clusters that are naturally cohesive and align well with the underlying concerns: 
\begin{enumerate}[leftmargin=*]
    \item \textbf{Directory Management (Orange):} \ttbreak{outputDirectoryStr}~(3), \ttbreak{inputDirectoryStr}~(4), \ttbreak{failedDirectoryStr}~(5), \ttbreak{createDir}~(24), \ttbreak{processDirectory}~(16), and \ttbreak{cpFile}~(17) are grouped together, reflecting their shared role in file system operations.
    
    \item \textbf{Program Initialization (Red):} \ttbreak{parseCommandLineParams}~(22), \ttbreak{validateParams}~(25), \ttbreak{schemaPathStr}~(6), and \ttbreak{main}~(9) form a cohesive cluster. Method~(22) parses command-line arguments, method~(25) validates the configuration, attribute~(6) is used by both, and method~(9) orchestrates the setup process.
    
    \item \textbf{Schema Processing (Blue):} \ttbreak{schema}~(7), \ttbreak{genTableSchema}~(10), and \ttbreak{sortAndParseLines}~(14) are grouped together, capturing their shared involvement in schema handling.
\end{enumerate}

Figure~\ref{fig:case-study-codebert} presents the embeddings generated by CodeBERT. It is clear that it forms a large cluster, at the bottom of the figure, containing 17 of the 25 entities, but our manual inspection of the source code reveals no clear semantic or structural rationale for aggregating these entities. 

Figure~\ref{fig:case-study-lsi} depicts the embeddings of LSI, which also forms a large cluster in the middle. It  aggregates most attributes into a single cluster, but fails to capture the relationships among methods. 
}

\subsubsection{Answer to Research Question} 
We now answer the research question: \emph{Does ConcernBERT outperform existing models in distinguishing distinct responsibilities among program entities, and how do clustering algorithm and distance metric choices affect this performance?} The answer is yes, ConcernBERT consistently and significantly outperforms all baseline models across all group sizes, from simple two-class scenarios to complex 100-class cases. This strong performance holds across a variety of clustering algorithms and distance metrics, demonstrating that its effectiveness is robust to configuration choices. Furthermore, the substantial performance gap between ConcernBERT and CodeBERT, despite their identical architecture, confirms that the advantage stems from its specialized, membership-based training process. Notably, ConcernBERT also outperforms more recent Transformer models such as GraphCodeBERT, CodeT5+, and ModernBERT, indicating that architectural advances alone do not capture concern-level semantics as effectively as ConcernBERT's training process.

\vspace{-0.2cm}
\section{Threats to Validity}
\label{sec:threats}
Here, we discuss threats to the validity~\cite{Wohlin2012} of our study.

\textit{External Threats to Validity.} In this study, we only examined \totalrepos Java projects. Its effectiveness on software written in other languages remains unknown and could vary due to differences in idioms, abstractions, and code structure. It is also possible that software projects in other communities or industrial projects may have domain-specific naming conventions that would require special attention. The evaluation method, constructing artificially shuffled classes, may not reflect the complexity of real software maintenance tasks. As such, ConcernBERT's high performance in this controlled setting may not translate into success in more realistic scenarios like tool-assisted refactoring or architectural recovery in production codebases.

\textit{Internal Threats to Validity.} We trained ConcernBERT on 1,978,071 files sourced from \traningrepos repositories. These files are assumed to represent cohesive classes, as our training approach relies on the assumption that members of the same class share similar concerns. However, not all classes in open-source repositories are equally cohesive. Some files, particularly in smaller or less mature projects, may exhibit low cohesion despite being included in the training set. \minor{Performing noise estimation and qualitative validation, such as manual sampling to assess class-level cohesiveness, and quantifying the impact of label noise on training and evaluation are future work.} Additionally, the model may require fine-tuning for projects in specific domains or communities.

\textit{Construct Threats to Validity.} To evaluate ConcernBERT, we created test instances (synthetic classes) by combining methods from different classes into one, and assessed how well each model can recover its original membership. But there is a gap between the evaluation method and what we actually intend to measure, that is, the model's ability to detect cohesive entities and responsibility boundaries in real-world scenarios. Synthesizing classes by arbitrarily combining methods from different source classes may not fully reflect the complexity and ambiguity of real design problems. Specifically, real low-cohesion classes typically result from gradual design erosion and may contain intertwined concerns and complex dependencies. In contrast, synthetic classes formed by randomly mixing unrelated methods create an artificially clear separation, potentially inflating ConcernBERT's performance. The task of restoring original membership assumes that class boundaries are well-defined and discrete, whereas real-world responsibilities often blur and overlap, especially in legacy or poorly maintained code.
\vspace{-0.3cm}
\section{Discussion}
We now discuss the implications of ConcernBERT for related research areas, and the directions for our future work. 

\textit{Accurate Cohesion Metrics.} Over the past few decades, many software metrics have been proposed to assess properties such as complexity~\cite{McCabe1976}, coupling~\cite{Chidamber1994, Briand1999,Poshyvanyk2008}, and cohesion~\cite{Chidamber1994,Briand1998, Marcus2005}. Among these, cohesion has proven especially difficult to define and measure, resulting in a wide range of competing metrics. A comprehensive review~\cite{Izadkhah2017} identified 32  cohesion metrics across structural, conceptual, and interface-based categories but highlighted the absence of rigorous benchmarks for systematic evaluation, leaving critical questions about cohesion assessment unresolved. Most recently, model-based cohesion metrics using semantic techniques have gained prominence, including the LSI-based C3~\cite{Marcus2005}, LCSM~\cite{Marcus2005}, and CLCOM5~\cite{Ujhazi2010}, the LDA-based MWE~\cite{Liu2009} and NT~\cite{Chen2017}, and the Doc2Vec-based COOC~\cite{Miholca2021} and LCOSM~\cite{Miholca2021}. However, these approaches are limited by the assumption that entities with similar token distributions share similar concerns. By learning from real examples of cohesive classes, ConcernBERT has the potential to significantly improve cohesion assessment, offering a more accurate foundation for future cohesion metrics.

\textit{Concern-driven Refactoring and Decomposition.}
Existing approaches to class decomposition and extract class refactoring~\cite{Lucia2008,Fokaefs2012,Bavota2013,Bavota2014,Akash2019,Alzahrani2021,Alzahrani2022,Lefever2025} typically combine structural and semantic information. Structural cues are derived from internal method calls or external client usages, while semantic information is obtained from models like LSI or LDA to identify cohesive subsets of entities~\cite{Bavota2013,Bavota2014,Akash2019,Lefever2025}. As our results demonstrate, ConcernBERT is significantly more effective at capturing the underlying concerns of program entities. Its strong performance on the Class-Membership Recovery Test, particularly in small-group scenarios, suggests that any of these class decomposition or extract class techniques could benefit from replacing their semantic component with ConcernBERT, resulting in more accurate and concern-aligned refactoring recommendations.

\textit{Better Technical Debt Detection and Presentation.}
With more effective cohesion metrics and improved complex class decomposition techniques, designers can more precisely monitor software evolution to proactively identify and prevent critical forms of technical debt, particularly god classes and spaghetti code. By operationalizing long-established design principles such as high cohesion and the single responsibility principle, ConcernBERT provides the foundation for actionable tooling, enabling teams to mitigate structural degradation and maintain sustainable software design.

\textit{Future Work.}
\revision{We acknowledge that, although our dataset is large in scale, it is limited to Java projects. Exploring more diverse datasets, as well as applying ConcernBERT to cluster large files written in other programming languages, is left for future work. In addition, a thorough exploration of the hyperparameter space is also our future work, as this study focuses on evaluating the effectiveness of the training strategy relative to the baselines.}

Our near-term future work will focus on developing new cohesion metrics and improving complex class decomposition techniques using ConcernBERT embeddings. We plan to collaborate with domain experts to refine these metrics and methods, incorporating expert feedback through targeted case studies. In addition, we will conduct industrial studies to rigorously validate the effectiveness of these approaches in real-world software development settings, particularly for monitoring software evolution and identifying or preventing technical debt.

\revision{We fully recognize that, although ConcernBERT produces better clusterings than the baseline models, identifying “single responsibility” in downstream tasks, such as god class decomposition, requires multiple factors beyond semantic similarity, such as internal and external dependencies and interactions among attributes, methods, and clients. We have begun exploring more sophisticated tasks such as architecture recovery and class extraction; however, a comprehensive and fair comparison would require a substantially different experimental setting, including practitioner-derived ground truth, and is therefore our future work.}

\vspace{-0.2cm}
\section{Conclusion}
In this paper, we introduced ConcernBERT, a novel embedding model designed to identify cohesive groups of program entities that collectively fulfill a common responsibility. \minor{ConcernBERT should be seen as a foundational representation for downstream analyses, such as refactoring, rather than as a complete, end-to-end refactoring solution.} \revision{ConcernBERT is trained with triplet loss at the entity (methods and attributes) that leverages class-membership context to learn responsibilities and concerns. We also contribute a large-scale replication dataset to support training and evaluation.} Through a thorough evaluation based on recovering original class memberships from randomly merged classes, we have shown that ConcernBERT consistently outperforms state-of-the-art models, demonstrating its ability to capture concern-level semantics. 

These findings support ConcernBERT's potential as a foundational concern-identification technique to support downstream tasks such as god class decomposition, extract class refactoring, and cohesion measurement, and take a significant step toward transforming long-standing principles like \textit{separation of concerns}, \textit{high cohesion}, and \textit{single responsibility} into actionable, automated techniques.

\section{Data Availability}
We provide all code, data, models, tables, and figures in a replication package at~\cite{Lefever2025a}.

\bibliography{main}

\end{document}